\documentstyle[pre,aps,preprint]{revtex}
\date{today}
\begin{document}
\title {Disordered Type-II Superconductors: A Universal Phase Diagram 
for Low-T$_c$ Systems\\
\vskip 1.25truecm
\normalsize
\noindent
S. S. Banerjee$^{1}$, A. K. Grover$^1$, M.J. Higgins$^{2}$, 
G. I. Menon$^{3}$\footnote{Corresponding Author, 
Email:menon@imsc.ernet.in},
P. K. Mishra$^4$, D. Pal$^{1}$, S. Ramakrishnan$^1$, 
T.V. C. Rao$^{4}$, G. Ravikumar$^4$, V. C. Sahni$^4$, 
S. Sarkar$^{1}$\footnote{Corresponding Author, Email:shampa@mailhost.tifr.res.in}
and C. V. Tomy$^{5}$
\vskip 0.1truecm
{\it
$^1$ Department of Condensed Matter Physics and Materials Science, Tata 
Institute of Fundamental Research, 
Mumbai-400005, 
India\\
$^2$ NEC Research Institute, 4 Independence
Way, Princeton, New Jersey 08540, U.S.A\\
$^3$The Institute of Mathematical Sciences, C.I.T Campus,
Taramani, Chennai 600 113, India\\
$^4$Technical Physics and Prototype Engineering Division, Bhabha Atomic 
Research Centre,
Mumbai-400085, India\\
$^5$ Department of Physics, Indian Institute of Technology, Powai, Mumbai-
400076, India}}
\date{\today}
\maketitle
\abstract 
{
A universal phase diagram for weakly pinned low-T$_c$ type-II
superconductors is revisited and extended with new proposals.  The
low-temperature ``Bragg glass'' phase is argued to transform first into
a disordered, glassy phase upon heating. This glassy phase, a
continuation of the high-field equilibrium vortex glass phase, then
melts at higher temperatures into a liquid. This proposal provides an
explanation for the anomalies observed in the peak effect regime of
2H-NbSe$_2$ and several other low-T$_c$ materials which is independent
of the microscopic mechanisms of superconductivity in these systems.
}
\vskip 0.5truecm
\noindent {\bf KEYWORDS:} Type-II superconductivity; Vortex state; Magnetic
phase diagram; Peak effect; Peak effect anomalies; Low T$_c$ systems

\newpage
\section{Introduction}
Random pinning destroys the long-ranged translational order of the
Abrikosov flux-line lattice (FLL)\cite{larkin,review1}.  This phase is
believed to be replaced by a new thermodynamic phase, the ``Bragg
glass'' phase, in which translational correlations decay as power
laws\cite{giam1}.  Experiments see a first-order, temperature-driven
melting transition out of the Bragg glass in weakly disordered single
crystals of the cuprate superconductors
Bi$_2$Sr$_2$CaCu$_2$O$_{8+\delta}$ (BSCCO) and
YBa$_2$Cu$_3$O$_{7-\delta}$ (YBCO), for fields $H$ less than a critical
value H$_{cr}$\cite{gammel,zeldov,ssten,khayko}.  For BSCCO, H$_{cr}
\simeq 500 G$; for YBCO H$_{cr} \simeq 10T$.

The Bragg glass (BG) transforms discontinuously \cite{khayko} into a
highly disordered {\em solid} phase as $H$ is increased at fixed low
temperature $T$\cite{giam1,zeldov,ssten,gaifullin}.  Such a disordered
phase, in which translational correlations decay
exponentially\cite{gingras}, is also encountered on temperature ($T$)
scans at $H > H_{cr}$; the data suggest a continuous transition into
this low-$T$ phase from the equilibrium disordered liquid at high
temperatures.  This low-temperature phase may be a thermodynamic
``vortex glass'' (VG) phase, analogous to a spin glass phase, separated
from the disordered liquid (DL) by a line of continuous phase
transitions\cite{fifi}. (Alternatively, it could be a ``frozen''
liquid, similar to a structural glass\cite{frozen} or even a highly
entangled phase\cite{ertaz,irrespective}).  Consensus phase diagrams
for disordered high-T$_c$ materials which incorporate these phases are
based principally on studies of the cuprates. Such phase diagrams show
the continuous VG-DL transition line meeting the first-order BG-DL and
BG-VG lines at a multicritical point
(T$_{cr}$,H$_{cr}$)\cite{gammel,zeldov,ertaz,vinokur,note}.

For the cuprates, properties such as intrinsic layering, anisotropy,
small coherence lengths and large penetration depths amplify the
effects of thermal fluctuations.  As a consequence, disordered phases
such as the flux liquid phase occupy much of the phase diagram; the
associated phase boundaries thus fall well below H$_{c2}$.  The
Ginzburg number $G_i$, (= $(k_BT_c/H_c^2\epsilon\xi^3)^2/2$), measures
the importance of thermal fluctuations: $\xi$ is the coherence length,
$\epsilon$ the mass anisotropy, $T_c$ the superconducting transition
temperature and $H_c$ the thermodynamic critical field\cite{review1}.
In most conventional low-T$_c$ materials $G_i \sim 10^{-8}$, whereas
$G_i \sim 10^{-2}$ for BSCCO and YBCO.  For superconductors with a
small G$_i$, phase boundaries such as the melting line lie very close
to H$_{c2}$ and are thus hard to assign without ambiguity.

In the low-T$_c$ system 2H-NbSe$_2$ (T$_c \simeq 7.2 $K), the effects
of thermal fluctuations are enhanced for reasons similar to those in
the cuprates.  For this material, G$_i \sim 10^{-4}$\cite{shobo}. For
the C15 Laves-phase superconductor CeRu$_2$ (T$_c \simeq 6.1 $K), G$_i
\sim 10^{-5}$, while for the ternary rare-earth stannides
Ca$_3$Rh$_4$Sn$_{13}$, (T$_c \simeq 8$K) and Yb$_3$Rh$_4$Sn$_{13}$,
(T$_c \simeq 7.6$K) , G$_i \sim 10^{-7}$.  For the quarternary
borocarbide YNi$_2$B$_2$C (T$_c \simeq 15.5$K), G$_i \sim 10^{-6}$.
Such relatively large values of G$_i$ imply that the phase diagram of
relatively clean, low-T$_c$ type-II superconductors can be studied
using such compounds, in regimes where the phases and phase transitions
discussed above in the context of the cuprates are experimentally
accessible.

These systems all exhibit a sharp ``Peak Effect'' (PE) in the critical
current density $j_c$, a measure of the force required to depin the
lattice\cite{belincourt,campbell}. The PE is the anomalous {\em
increase} in j$_c$ seen  close to the $H_{c2}(T)$; this increase
terminates in a peak, as $H$ (or $T$) is increased, before $j_c$
collapses rapidly in the vicinity of H$_{c2}$.  One mechanism for the
peak effect \cite{pippard} is the softening of the shear elastic
modulus of the vortex solid as $H$ approaches H$_{c2}(T)$. Lattice
(shear) distortions which maximize the energy gain due to the pinning
thus cost less elastic energy. As a consequence, j$_c$ increases since
the flux-line lattice can adapt better to its pinning environment.  In
an approximation believed to be valid for weakly pinned vortex line
systems, $j_cB=(\frac{n_p<f_p^2>}{V_c})^{\frac12}$ \cite{lo}.  Here $B$
is the magnetic induction, n$_p$ the density of pins, f$_p$ the
elementary pinning interaction and V$_c$ the correlation volume of a
Larkin domain. V$_c$ and $f_p$ are both suppressed as $H \rightarrow
H_{c2}$.  A peak in $j_c$ thus implies a more rapid reduction in V$_c$
as compared to $<f_p^2>$. Above the peak, the rapid collapse in the
shielding response\cite{explan} is governed entirely by the suppression
of pinning on approaching $H_{c2}$.

A peak effect can also occur across a melting transition, when a shear
modulus vanishes abruptly\cite{shobo}; early studies of the peak effect
in 2H-NbSe$_2$ suggested such a mechanism for the peak effect in this
material.  Evidence in favour of this correlation between melting
phenomena and (some but not all) sharp peak effects has mounted in
recent years.

The locus of points in ($H,T$) space from which $j_c$ begins to increase to its
peak value defines a line called T$_{pl}(H)$; the locus of maxima of
$j_c$ is T$ _p(H)$\cite{shobo}. The regime of fields and temperatures
between T$_{pl}$ and T$_p$ is the peak regime. This regime is
dynamically anomalous.  Its properties (the PE anomalies) include the
following: (i)  A profound history dependence of static and dynamic
response\cite{shobo,steingart,kes,henderson,satya1,satya11,satya4,shampa},
(ii) large noise signals in electrical transport and ac susceptibility
measurements \cite{shobo,satya1,marley,merithew}, (iii) substantial
memory effects\cite{shobo,henderson,henderson1,xiao}, (iv)
``switching'' phenomena by which the system can transit between j$_c$
values characteristic of a field cooled (FC) or zero-field-cooled (ZFC)
state on applying $T$ or ac amplitude pulses\cite{satya4} and (v)
``open hysteresis loops'' in thermal cycling \cite{satya1}. Why these
anomalies arise is still incompletely understood; an explanation is
outlined in this paper.

This paper presents a phase diagram (Fig. 1) for weakly pinned
low-T$_c$ type-II superconductors with point pinning
disorder\cite{caveat}, which incorporates and extends earlier related
proposals\cite{caveat0}.  We propose the following:  The equilibrium
high-field VG phase survives in the intermediate field regime,
$\Phi_0/\lambda^2 < H < H_{cr}$\cite{hcr} as a sliver intervening
between quasi-ordered and liquid phases, as shown in Fig. 1.  This
sliver is precisely the regime in which PE anomalies -- non-trivial
relaxation behaviour, memory effects and substantial history dependence
in measured properties -- are seen.  We propose that such properties
are generic and arise due to the continuation of the high field glassy
behaviour into a regime where sensitivity to external perturbations is
enhanced and relaxation times become accessible. The inset to Fig. 1
expands the regime of fields and temperatures shown in the boxed region
of the main figure and depicts the expected phase behaviour at {\em
low} fields (H$_{c1} < H < \Phi_0/\lambda^2$) , where the BG-VG phase
boundary shows reentrant behaviour (see below).

Our phase diagram differs from the conventionally accepted one; it
contains no multicritical point and the BG phase never melts directly
into the liquid.  We believe that this is due to the substantial
thermal renormalization of disorder relevant to the high-T$_c$ cuprates
close to the melting transition; such effects suppress the
irreversibility line\cite{ssten} to below the melting line and can
effectively render the sliver phase
unresolvable\cite{pal1,menon,notehere}.

Our proposals are based on a study of the systematics of the peak
effect in a variety of low T$_c$  systems {\it via} ac susceptibility
and dc magnetization measurements on relatively pure single
crystals\cite{satya1,satya11,satya10,shampa2,pal2}.  AC susceptibility
measurements access j$_c$ via the real part of the ac susceptibility
$\chi '$ in the following way:  Once the applied ac field penetrates
the sample fully, $\chi '\sim -\beta\frac{J_c }{h_{ac}}$, from the
Bean's critical state model\cite{jsps,bean}.  Here $\beta$ is a
geometry and size dependent factor, while $h_{ac}$ is the magnitude of
the applied ac field. At very low temperatures, $\chi ' \approx -1$,
indicating perfect screening of applied fields.  Non-monotonic
behaviour in $\chi '$ reflects non-monotonicity in $j_c(H, T)$; the
minimum in $\chi '$ corresponds to a peak in $j_c$.

\section{Experimental Results} Results of typical measurements of $\chi
'$ performed on 2H-NbSe$_2$\cite{satya1}, CeRu$_2$\cite{satya1},
Ca$_3$Rh$_4$Sn$_{13}$ \cite{shampa}, Yb$_3$Rh$_4$Sn$_{13}$
\cite{shampa2} and YNi$_2$B$_2$C \cite{pal2} are shown in Figs.
2(a)-(e) for field values as indicated in the figures and at
temperatures close to T$_c(H)$. The two curves shown in each figure
refer to different thermomagnetic histories -- zero-field-cooled (ZFC)
and field-cooled (FC).  The structure in $\chi '$ is broadly similar
for both ZFC and FC histories, although FC histories yield far less
dramatic peak effects.  Field cooling in these samples generically leads to 
higher j$_c$ values (and thus smaller correlation volumes \cite{lo})
than cooling in zero field. This is in contrast to the phenomenology of 
spin glasses. Such behaviour is also opposite to that observed for
high-T$_c$ materials.  An explanation for this anomalous behaviour will
be presented in what follows.

Note the existence of abrupt discontinuities in $\chi '$ in Figs.
2(a)-(e).  These data show discontinuities at T$_p$ and T$_{pl}$, which
have the following properties:  (i) their locations are independent of
the ac field amplitude and the frequency over a substantial range, (ii)
two discontinuities associated with the onset and the peak positions of
the PE are generically seen, although more complicated intermediate
structure is manifest in some samples (iii) their locations correlate
precisely with $T_p$ and $T_{pl}$ measured in transport measurements --
such measurements show that the {\em locations} of such features are
independent of the magnitude of the driving current and are hence
properties of the sample and not of the measurement technique and (iv)
they are seen both in ZFC and FC data {\em and at identical locations}.
These properties indicate that it is thermodynamic behaviour which is
being reflected and that these phenomena are independent of the
measurement technique.

As $j_c$ collapses from its peak value above T$_p$, the diamagnetic
$\chi '$ response becomes a paramagnetic one across the irreversibility
temperature $T_{irr}$ \cite{satya11}. Above $T_{irr}$, $j_c \approx 0$
and the mixed state of the superconductor has reversible magnetic
properties. In the reversible phase, differential magnetic response is
positive ({\it i.e.}, diamagnetic dc magnetization decreases as $H$ or
$T$ is increased).  Such a differential paramagnetic effect identifies
the depinned ($j_c$$=0$) state.  A $\chi '(T)$ measurement at fixed $H$
thus yields three characteristic temperatures, denoted as
$T_{pl}$,~T$_p$, and ~T$_{irr}$ \cite{satya11}.

Fig. 3(a) shows the peak onset at (H$_{pl}$,T$_{pl}$), the location of
the peak at (H${_p}$,T${_p}$), and the apparent irreversibility line
(H$_{irr}$,T$_{irr}$), in 2H-NbSe$_2$; the behaviour in the
intermediate field regime is qualitively the same for CeRu$_2$,
Ca$_3$Rh$_4$Sn$_{13}$, Yb$_3$Rh$_4$Sn$_{13}$ and YNi$_2$B$_2$C (cf.
Figs. 3(b) to 3(e)). The T$_p$ and T$_{pl}$ lines begin to separate at
high values of the field; at low field values, a similiar broadening of
the peak regime also occurs.  The peak effect is sharpest at
intermediate field values.  At high and low $H$, the discontinuities
become weaker.  In some circumstances the discontinuities evolve into a
broad dip in $\chi '$(T) and resistive response\cite{satya13}.  For
sufficiently low fields in 2H-NbSe$_2$, the T$_p$ line curves
backwards, moving to lower field values as the temperature is
decreased. We have argued elsewhere that this behaviour signals the
reentrant character of the low-field melting transition in the pure
system\cite{reentrant}.

The nature of the phase boundaries at low values of the field can also
be studied through j$_c$(H)/j$_c$(0) vs H/H$_{c2}$ plots at fixed
$T$\cite{satya12}.  Fig. 4 shows such plots for a single crystal of
2H-NbSe$_2$ at different values of t($\approx$T/T$_c$).  At t=0.75, the
power law behaviour in j$_c$(H) marks the collectively pinned  regime,
which we identify as a Bragg glass. This power law regime terminates
for large $H/H_{c2}$, when $j_c(H)$ begins to vary anomalously,
signalling the PE regime. For low fields, another anomalous variation
in j$_c$(H) (see Fig. 4) marks the transition from the collectively
pinned power-law regime to another regime, which some of us have
identified in previous work as a  ``small bundle pinning''
regime\cite{satya12}. This anomalous variation has been termed a
``Plateau Effect'' by some of us \cite{satya12}.

Recent transport measurements of Paltiel {\it et al} \cite{paltiel0} in
single crystals of 2H-NbSe$_2$ confirm that the onset of each of these
anomalies in j$_c$(H) (both at low and high field ends) occurs
abruptly.  The anomalous variation in j$_c$ thus exhibits reentrant
behaviour\cite{satya11,reentrant,satya13,satya12} as H is varied.
Fig.  5 shows the boundary which separates the collectively pinned
power-law regime from the low and high field regimes in which j$_c$(H)
varies anomalously.  The power law region shrinks as the temperature
increases. Above a characteristic temperature (which correlates with
the quenched random disorder in the crystal\cite{jsps,satya12}), this
boundary is not encountered in an isothermal scan. This is reflected in
the monotonic variation of j$_c$(H) at t=0.983 shown in Fig. 4.

\section{Discussion}
An interpretation of the above mentioned phenomena is now proposed; it 
builds on the work of Refs.\cite{satya1}, \cite{shampa} and \cite{satya12} 
by emphasizing the relation to the equilibrium phase diagram of Fig. 1.
The presence of two discontinuities in $\chi '$ {\it vs.} $T$ in Figs.
2(a)-(e) implies that three different phases are encountered in $T$
scans which commence from within the Bragg glass at intermediate $H$.
If the second discontinuity is associated to the transition into a
uniform fluid, the first must be interpreted as a transition between a
relatively ordered phase and a more disordered phase.  The melting of
the Bragg glass on a temperature scan is thus argued to be a two-stage
process, with the transformation into the liquid preceded by a
transformation into an intermediate state with a larger
j$_c$\cite{satya1}.  The intermediate state defines a small ``sliver''
in the phase diagram; its width in temperatures is bounded by T$_p$ and
T$_{pl}$.

What is the fate of the sliver (or PE regime) at high fields?. We argue
here that the sliver regime {\em must} broaden out into the {\em equilibrium
vortex glass} phase expected at low temperatures and high fields; these
phases are smoothly connected, as shown in Fig. 1.  This phenomenology,
as well as the absence of a multicritical point, is suggested both by
the increasing separation of the T$_p$ and T$_{pl}$ lines at high
fields and the strongly anomalous behaviour seen in the peak regime.
The phase diagram of Fig. 1 immediately suggests a simple physical
origin of PE anomalies: {\em the peak regime is anomalous precisely
because it is glassy}.  In terms of Fig. 1, a scan in $T$ first
encounters the BG-VG phase boundary.  On further increasing $T$, the
VG-DL phase boundary is encountered, whereupon the second transition
occurs.  Experimentally, nowhere in the intermediate field regime do
the two discontinuities appear to merge.  Thus, the
``sliver'' of intermediate phase appears to intrude smoothly between
the high-field and the low field end, as corroborated by Fig. 5.
At low fields, the data suggest the phase boundaries shown in the
inset of Fig. 1, in particular the reentrant nature of the BG-VG
phase boundary\cite{notabene}. 

Of the two transitions out of the Bragg glass phase at intermediate
fields, the second -- between VG and DL phases -- can be argued to be
the true remnant of the underlying first-order melting transition in
the pure flux line lattice system (see below).  Thermodynamic
signatures of the underlying (discontinuous) melting transition in the
pure system should thus be strongest across this boundary; this is
consistent with magnetization experiments on untwinned and twinned
single crystals of YBCO\cite{ssten,pal1,nishizaki,ishida,shi}.
However, the two stage nature of the transition implies that some part
of the difference in density between ordered and liquid phases can be
accomodated across the width of the sliver regime\cite{nravi}. 

We argue that the variation in $\chi '$ between the ZFC and the FC
histories simply reflects the fact that the system, on field cooling,
first encounters the irreversibility line and then the highly
disordered glassy sliver phase. Since it is in the irreversible region,
it is trapped into a metastable state easily and only drastic
mechanical perturbations, such as a ``shaking'' of the crystal via
large amplitude ac fields, enable it to lower its free
energy\cite{satya4}.  In  contrast, zero-field cooling can ensure that
the thermodynamically stable Bragg glass phase is the preferred state,
since the intermediate irreversible glassy state is not encountered.

The anomalies in the noise spectrum measured in the PE regime are
explained in the following way:  If the VG phase is a genuine glass (in
the sense that any appropriately coarse-grained free energy landscape
has many metastable minima, {\it i.e.}, it is {\em complex}),
transitions between such mimima should yield non-trivial signals in the
noise spectrum.  This is consistent with the suggestive observations of
Merithew {\it et al.} \cite{merithew}, who find that the noise arises
from {\em rearrangements} of the {\em pinned} condensate in the peak
regime. This implies that the noise originates in the static properties
of the underlying phase and not exclusively from its flow
properties\cite{zhang}.  We emphasize that the central idea which
differentiates our proposals from those of others is our connection of
these dynamical anomalies to the static phase diagram proposed in 
Fig. 1.

Paltiel {\it et al.}\cite{paltiel} have proposed that many features of
the PE regime can be understood in terms of the surface barriers to
vortex entry. In an imaginative scenario, these authors suggest that
vortices enter in a disordered state at the boundaries and then anneal
in the (ordered) bulk. It is argued that the disordered phase invades
the bulk from the boundaries as the peak regime is entered, leading to
slow relaxation and memory effects.

The ideas presented here substantiate and extend these proposals by
linking them to the underlying thermodynamic behaviour. We have
suggested that the bulk is highly disordered in the peak regime as a
consequence of the connection of this regime to the VG phase. This link
is embodied in the {\em static} phase diagram of Fig. 1 and is {\em
independent} of the details of the dynamics.  We can thus argue that
vortices injected from the boundaries in the PE regime equilibrate
slowly because the structure to which they are equilibrating is a glass
with many metastable states.  This extension of the picture of Paltiel
{\it et al.} \cite{paltiel} has the following appealing features:
(i) it rationalizes the observation of discontinuous changes in $\chi
'$ {\it vs.} $T$; such discontinuous behaviour cannot be otherwise
obtained, (ii) it demonstrates unambiguously why such behaviour is
unique to the PE regime, and (iii) it clarifies the connection of PE
behaviour with the underlying bulk thermodynamic melting transition in
the pure system.

Recent simulations also support the ideas presented here. A. van
Otterlo {\it et al.}\cite{otterlo} remark that the existence of a
sliver of the VG phase preempting a direct BG-DL transition is a
possibility consistent with the simulation results.  Other simulations
suggest that the experimental observation of a split fishtail peak in a
relatively pure untwinned YBCO sample\cite{deligiannis} reflects a
sequence of two transitions out of the Bragg glass
phase\cite{otterlo1}.  These proposals have their precise counterpart
in Fig. 1;  as argued here and
elsewhere\cite{satya1,shampa,satya10,pal3}, the peak effect in
2H-NbSe$_2$ and related materials actually comprises two separate
transitions -- the BG-VG and the VG-DL transitions. The phase diagram
of Fig. 1 illustrates how the peak effect in $T$ scans at intermediate
values of the field should evolve smoothly and continuously into
behaviour characteristic of the BG-VG field driven transition both at
low and high fields.  This is a simple consequence of the absence of an
intervening multicritical point in the phase diagram of Fig. 1. In
other words, the robustness of the two-step character of the PE anomaly
which occurs as a precursor to H$_{c2}$ rules out the possibility of a
multicritical point.

We emphasize that the putative multicritical point and the single sharp
thermal melting transition featured in theoretical models of the phase
behaviour of high-T$_c$ materials\cite{gammel,vinokur} are not mandated
by theory.  Indeed, our proposal is simpler and possibly more generic.
The instability of the Bragg glass phase to a spontaneous proliferation
of a high density of dislocations could well be preempted by a
transition into a phase with an intermediate density of dislocations;
nothing in the theory forbids this.

What might be the structure of the sliver phase?  We suggest that it is
a ``multi domain'' structure, with fairly well-ordered crystalline
domains and local amorphous or liquid-like regions.  Translational (and
orientational) correlations would then decay strongly beyond the
typical domain size. In this picture, the second jump in $\chi '$ is
related to the melting transition of the multi domain
solid\cite{disprl,tang}, while the first represents the ``fracturing''
\cite{satya1} of the BG phase into such a multi domain structure.
Provided the typical size of domains is much larger than the
typical correlation length at the melting transition, one expects sharp
signals of melting.  These signals would progressively decrease in
amplitude as the system became more disordered, or, given the expected
correlation between $H$ and disorder, as $H$ is increased.

The above interpretation is consistent with the intermediate jumps in
$\chi'$ seen in some measurements of the peak effect (cf. Fig. 2(e) for
Yb$_3$Rh$_4$Sn$_{13}$). These jumps could then be taken as signalling
the melting of particularly large domains or perhaps a
percolation of liquid-like or amorphous regions.  Such ``precursor''
melting could arise due to disorder-induced variations in melting
temperatures in different regions of the sample.  This physical picture
agrees with the proposals of Paltiel {\it et al.}\cite{paltiel} who
argue that the PE regime reflects the dynamic coexistence of ordered
and disordered phases. (We argue here that such behaviour in the peak
regime actually reflects the complex nature of the statics).  The
observation of open hysteresis loops in experiments on 2H-NbSe$_2$ and
other low T$_c$ systems and their rationalization in terms of a {\em
fracturing} of the disordered PE state\cite{satya1,shampa} point to
this interpretation. Such fracturing would then be a consequence of
large free energy barriers separating such a multi domain or
``microcrystalline'' state, with its relatively short-ranged
translational correlations, from the BG phase, where such correlations
are infinite ranged. These ideas also correlate well with the
observation of a critical point of the vortex glass-liquid transition
in the experiments of Nishizaki {\it et al.} \cite{nishizaki} on YBCO,
beyond which the distinction between these phases vanishes.

Voltage noise measured in the flowing state created just above the
depinning threshhold shows a marked increase in the peak regime of
2H-NbSe$_2$.  Marley {\it et al.} \cite{marley} have probed such low 
frequency, broadband noise through transport measurements and found 
strong dependences on the current and $H$ in the power, spectral 
shape and non-gaussian character of the noise. Related enhancements
in the $\chi '$ noise have been observed by Banerjee 
{\it et al.}\cite{satya1}. Merithew {\it et  al.}  \cite{merithew} 
have measured the higher order fluctuators of
this noise to extract the number of independent channels or
contributors to the noise spectrum. A fairly small value for the number
of correlators is derived, justifying its non-gaussian
character\cite{merithew}. Such a small number of correlators emerges
naturally in terms of a picture of large, correlated solid-like ``chunks'', 
as in a multi domain arrangement, with each chunk fluctuating
collectively.  Recent simulations are also consistent with this 
possibility\cite{rudnev}.  Direct tests of this proposal, through 
spatially resolved Hall probe measurements, Bitter decoration or 
neutron scattering would be welcome.

In conclusion, this paper proposes a general phase diagram for weakly
pinned type-II superconductors based on a study of the peak effect
systematics in a variety of low-T$_c$ superconductors. The arguments
presented here do not rely on the particular microscopic mechanism of
superconductivity in these materials and are hence general.  We have
argued that the sensitivity to external perturbations, novel history
effects and switching phenomena exhibited in the peak regime derive
from the {\em static} complexity of the vortex glass state.  Such
complexity is envisaged in theories of the low-temperature properties
of structural and spin glasses.  The PE regime may provide a
remarkable new testing ground for these theories. Further work is
clearly called for to test these ideas. More results, in particular on
issues related to the phase diagram of high-T$_c$ materials will be
published elsewhere\cite{pal1,menon}.

\noindent{\bf Acknowledgements}

We thank Mahesh Chandran, Chandan  Dasgupta, Deepak Dhar and Nandini Tridevi
for useful discussions.  In particular, we are deeply indebted to 
Shobo Bhattacharya for collaborations in the experimental studies and for
generously sharing many critical insights with us. We would also like to 
thank him for a critical reading of this manuscript.

\begin{figure}
\caption
{
Proposed phase diagram for a low-T$_c$ Type-II superconductor
incorporating the effects of thermal fluctuations and quenched random
point pinning. For a description of the phases, see the text. Note that
the vortex glass phase intrudes between Bragg glass and disordered
liquid phases everywhere in the phase diagram. The disordered
liquid may have irreversible or reversible magnetic
properties. In the intermediate
field range, the glass phase is confined to a slim sliver but broadens
out again for sufficiently low fields. The irreversibility line (dotted)
lies above the VG-DL phase boundary (see text). The inset to the figure
expands the boxed region shown in the main panel at low fields and 
temperature values close to T$_c(0)$ and illustrates the reentrant
nature of the BG-VG phase boundary at low fields. 
}
\label{Fig1}
\end{figure}

\begin{figure}
\caption{
Temperature dependence of $\chi '$ for vortex arrays created in zero
field cooled (ZFC) and field cooled (FC) modes at the fields indicated
in single crystals of (a) 2H-NbSe$_2$,(b) CeRu$_2$, 
(c) Ca$_3$Rh$_4$Sn$_{13}$, (d) YNi$_2$B$_2$C  and (e) Yb$_3$Rh$_4$Sn$_{13}$.
Note the two sharp changes in $\chi '$ response
at the onset temperature $T_{pl}$ and peak temperature $T_p$ in the ZFC
mode. The difference in $\chi '$ behavior between ZFC and FC histories
disappears above the peak temperature.
}
\label{Fig2}
\end{figure}
\begin{figure}
\caption
{
Vortex phase diagrams (for $H > 1$ kOe) of (a) 2H-NbSe$_2$,
(b) CeRu$_2$, (c) Ca$_3$Rh$_4$Sn$_{13}$, (d) YNi$_2$B$_2$C and 
(e) Yb$_3$Rh$_4$Sn$_{13}$.  Note that
the lines marking the onset of the PE (H$_{pl}$), the onset of the
reversibility (H$_{irr}$) and the upper critical field (H$_{c2}$) can
be distinctly identified.  For a description of the different phases
shown in this figure, see text.
}
\label{Fig3}
\end{figure}
\begin{figure}
\caption
{Plot of log(j$_c$/j$_c$(0)) vs log(H/H$_{c2}$) in a crystal of 
2H-NbSe$_2$ at three different temperatures. The power law region
at t=0.75 has been marked and the two regions of anomalous variation in
j$_c$(H) have been identified as the Peak Effect and the Plateau Effect,
following Ref. 46 (see text for details).
}
\label{Fig4}
\end{figure}
\begin{figure}
\caption
{The (H,T) phase diagram showing the demarcation of the collectively
pinned region (corresponding to the power law behaviour in j$_c$(H)) 
as distinct from the regions corresponding to anomalous variation (i. e.,
the Peak Effect and the Plateau Effect) in j$_c$(H) in 2H-NbSe$_2$.
Note that the collective pinned region is sandwiched between
region of anomalous variation in j$_c$(H) at high as well as
at low field ends.
}
\label{Fig5}
\end{figure}


\begin{thebibliography}{99}
\bibitem{larkin} A.I. Larkin, {\it Zh. Eksp. Teor. Fiz.}, { 58}
(1970) 1466 [{\it Sov. Phys. JETP} { 31} (1970) 784].
\bibitem{review1} G. Blatter, M. V. Feigelman, V. B. Geshkenbein,
A. I. Larkin and V. M. Vinokur, {\it Rev. Mod. Phys.} { 66} (1994) 1125.
\bibitem{giam1} T. Giamarchi and P. Le Doussal, {\it Phys. Rev. Lett.}
{ 72} (1994) 1530; T.  Natterman, {\it Phys. Rev. Lett.} { 64}
(1990) 2454.  
\bibitem{gammel} See P. L. Gammel, D. A. Huse and D. J. Bishop, in
{\it Spin Glasses and Random Fields}, ed. A. P. Young, World
Scientific, Singapore (1998).
\bibitem{zeldov} E. Zeldov, D. Majer, M. Konczykowski, 
V.B. Geshkenbein and V.M. Vinokur, {\it Nature} { 365} (1995) 375.
\bibitem{ssten} T. Nishizaki and N. Kobayashi, {\it Supercon. Sci. 
Tech.} { 13} (2000) 1.
\bibitem{khayko} B. Khaykovich, M. Konczykowski, E. Zeldov, R. A
Doyle, D. Majer, P. H. Kes and T. W. Li, {\it Phys. Rev. B}
{ 56} (1997) R517.
\bibitem{gaifullin} M. B. Gaifullin, Y. Matsuda, N. Chikumoto,
J. Shimoyama and K. Kishio, {\it Phys. Rev. Lett.} { 84} (2000) 2945;
C. J. van der Beek, S. Colson, M. V. Indenbom and M. Konczykowski,
 {\it Phys. Rev. Lett.} 84 (2000) 4196; Y. Nonomura 
and X. Hu, ({\it cond-mat/0002263}).
\bibitem{gingras} M. J. P. Gingras and D. A. Huse, {\it Phys. Rev. B}
{ 53} (1996) 15193.
\bibitem{fifi} D. S. Fisher, M. P. A. Fisher and D. A. Huse, {\it Phys. Rev. 
B} { 43} (1991) 130. 
\bibitem{frozen} G. I. Menon, C. Dasgupta and 
T.V. Ramakrishnan, {\it Phys. Rev. B} { 60} (1999) 7607.
\bibitem{ertaz} D. Ertaz and D. R. Nelson, {\it Physica C} { 272}
(1996) 79.
\bibitem{irrespective} Irrespective of the nomenclature, experiments
agree that this high-field low-temperature phase is glassy in nature. 
There is experimental evidence that the DL-VG transition is accompanied 
by a substantial increase (quite possibly a divergence) of relaxation 
times (see Ref. \cite{gammel}) although there is no consensus as to 
whether a description in terms of a single universal set of 
exponents is possible (see e.g. S. Misat, P.J. King, R.P. Campion, P.S.
Czerwinka, D. Fuchs and J.C. Villegier, {\it J. Low Temp. Phys},
{ 117} (1999) 1381). Pronounced history dependence is also seen in the
VG phase, see e.g. S. Kokkaliaris, A.A. Zhukov, S.N. Gordeev, P.A.J. de Groot,
R. Gagnon and L. Taileffer {\it J. Low Temp. Phys.}, { 117}
(1999) 1341; A.P. Rassau, S.N. Gordeev, P.A.J. de Groot, R. Gagnon and 
L. Taileffer, {\it Physica B}, {284-288} (2000) 693; M. Konczykowski,
S. Colson, C.J. van der Beek, M.V. Indenbom, P.H. Kes and E. Zeldov,
{\it Physica C} { 332} (2000) 219; D. Giller, A. Shaulov, R. Prozorov,
Y. Abulafia, Y. Wolfus, L. Burlachkov, Y. Yeshurun, E. Zeldov, V.M.
Vinokur, J.L. Peng and R.L. Greene, {\it Phys. Rev. Lett}, { 79},
(1997) 2542.
\bibitem{vinokur} V. Vinokur, B. Khaykovich, E. Zeldov, M.
Konczykowski, R. A. Doyle and P. H. Kes, {\it Physica C}
{ 295} (1998) 209.
\bibitem{note} There is no consensus on the nature of the putative
multicritical point. If, as recently suggested, the BG-VG transition
is first order, the point at which the (presumably continuous) VG-DL 
transition line meets the BG-VG phase boundary should be a critical 
end-point. We argue here that a multicritical point may not be
generic.
\bibitem{shobo} S. Bhattacharya and M. J. Higgins, 
{\it Phys. Rev. B} { 49} (1994) 10005; 
{\it Phys.  Rev. B} { 52} (1995) 64, 
{\it Phys. Rev. Lett} { 70} (1993) 2617. See, also, M. J. Higgins and S. 
Bhattacharya, Physica C { 257} (1996) 232 and references cited therein. 
\bibitem{belincourt} T. G. Belincourt, R. R. Hake and D. H. Leslie,
{\it Phys. Rev. Lett.} { 6} (1961) 671.
\bibitem{campbell} A. M. Campbell and J. E. Evetts, {\it Advan. Phys.} 
{ 21} (1972) 199.
\bibitem{pippard} A. B. Pippard, {\it Philos. Mag.} { 19} (1969) 217.
\bibitem{lo} A. I. Larkin and Yu. N. Ovchinnikov, 
{\it Sov. Phys. JETP} { 38} (1974) 854; 
{\it J. Low Temp. Phys.} { 34}  (1979) 409. 
\bibitem{explan} As measured, for example, through the real part
($\chi '$) of the complex ac susceptibility.
\bibitem {steingart}  M. Steingart, A. G. Putz and E. J. Kramer, 
J. Appl. Phys. { 44} (1973) 5580.
\bibitem{kes} R. Wordenweber,  P. H. Kes and C. C. Tsuei,  Phys. 
Rev. B {33}(1986) 3172. 
\bibitem{henderson}  W. Henderson, E. Y. Andrei, M. J. Higgins and
S. Bhattacharya, {\it Phys. Rev. Lett.} { 77}  (1996) 2077.
\bibitem{satya1} S. S. Banerjee, N.G. Patil, S. Saha, S. Ramakrishnan,
A.K. Grover, S. Bhattacharya, G. Ravikumar, P.K. Mishra, T. V. C.
Rao, V. C. Sahni, M. J. Higgins, E. Yamamoto, Y. Haga, M. Hedo, Y. Inada and
Y. Onuki, {\it Phys. Rev. B} { 58}, (1998) 995.
\bibitem{satya11} S. S. Banerjee, S. Saha, N. G. Patil, S. Ramakrishnan,
A. K. Grover, S. Bhattacharya, G. Ravikumar, P. K. Mishra, T. V. C. Rao,
V. C. Sahni, C. V. Tomy, G. Balakrishnan, D. Mck. Paul, M. J.Higgins,
Physica C { 308}  (1998) 25.
\bibitem{satya4} S. S. Banerjee, N.G. Patil, S. Ramakrishnan, A.K.
Grover, S. Bhattacharya, G. Ravikumar, P. K. Mishra, T. V. C. Rao, 
V. C. Sahni, M. J. Higgins {\it Appl. Phys. Lett.} { 74} (1999) 126.
\bibitem{shampa} S. Sarkar, D. Pal, S.S. Banerjee, S. Ramakrishnan,
A.K. Grover, C. V. Tomy, G. Ravikumar, P. K. Mishra, V. C. Sahni, 
G. Balakrishnan, D. McK Paul and S. Bhattacharya, {\it Phys. Rev. B}
 61 (2000) 12394.
\bibitem{marley} A. C. Marley, M. J. Higgins and S. Bhattacharya,
{\it Phys. Rev. Lett.} { 74} (1995) 3029.
\bibitem{merithew} R. D. Merithew, M.W. Rabin, M. B. Weissman, M. J. 
Higgins and S. Bhattacharya, {\it Phys. Rev. Lett.} { 77}  
(1996) 3197.
\bibitem{henderson1} W. Henderson, E. Y. Andrei and M. J. Higgins,
{\it Phys. Rev. Lett.} { 81} (1998) 2352.
\bibitem{xiao} Z. L. Xiao, E. Y. Andrei and M. J. Higgins, {\it Phys.
Rev. Lett.} { 83} (1999) 1664.
\bibitem{caveat} For extended disorder such as twin boundaries, 
a ``Bose-glass'' description of the phases and phase transitions 
may be more appropriate.  See D. R. Nelson and V. M. Vinokur, 
{\it Phys. Rev. B} { 48} (1993) 13060.
\bibitem{caveat0} Phase diagrams closely related to the one exhibited
in Fig. 1 have been proposed earlier; see {\it e.g.}. 
Ref.\cite{satya11,satya10,jsps}. Fig. 1 differs from these in the
following respects: (i) the different 
phases (Bragg Glass, Vortex Glass and Disordered Liquid) are completely 
enclosed from each other by transition boundaries as shown; Fig. 1 
{\em explicitly} rules out a multicritical point (or critical end 
point) whereas earlier phase diagrams left this possibility open,
(ii) it includes only those phases and phase transition lines which 
we believe can be justified in equilibrium and (iii) the word
``plastic glass'' which carries a dynamic connotation and was used
in earlier work to describe the PE regime, is replaced by ``vortex
glass'' to indicate our understanding that the measurements, though
dynamic in nature, reveal information about the static phases.
\bibitem{satya10} S. S. Banerjee, N. G. Patil, S. Ramakrishnan,
A. K. Grover, S. Bhattacharya, G. Ravikumar, P. K. Mishra, T. V. C. Rao,
V. C. Sahni, M. J. Higgins, C. V. Tomy, G. Balakrishnan and D. McK Paul,
{\it  Phys. Rev. B} { 59}  (1999) 6043.
\bibitem{jsps} S. S. Banerjee, S. Ramakrishnan, D. Pal, S. Sarkar,
A. K. Grover, G. Ravikumar, P. K. Mishra, T. V. C. Rao, V. C. Sahni,
C. V. Tomy, M. J. Higgins and S. Bhattacharya, {\it J. Phys.
Soc. Jpn. Suppl.} 69 (2000) 262.
\bibitem{hcr} For low-T$_c$ systems, H$_{cr}$ and T$_{cr}$ notionally
represent the point in the H-T phase diagram where broadening of
the narrow sliver regime at high fields becomes significant.
\bibitem{pal1} D. Pal {\it et al.}, {\it unpublished}.
\bibitem{menon} G. I. Menon, {\it unpublished}.
\bibitem{notehere} For very pure samples of low-T$_c$ 
materials (such as sample A of 2H-NbSe$_2$ referred to in 
Ref. 25), the T$_{pl}$ and T$_p$ lines cannot be
resolved separately, although an extremely narrow peak effect 
(its width is smaller than the width of the zero-field
superconducting phase transition) is seen. In more disordered 
samples such as the one discussed here, the two-step nature 
of the transition is clearly evident.  
\bibitem{shampa2} S. Sarkar, S.S. Banerjee, A.K. Grover, S. Ramakrishnan, S. Bhattacharya,
G. Ravikumar, P. K. Mishra, V. C. Sahni, C. V. Tomy, G. Balakrishnan, D. McK Paul and 
M. Higgins, {\it Proceedings of the M2S-HTSC VI}
held at Houston, U.S.A, Feb. 2000, Physica C (in press).
\bibitem{pal2} D. Pal, Shampa Sarkar, S. S. Banerjee, S. Ramakrishnan, 
A. K. Grover, S. Bhattacharya, G. Ravikumar, P. K. Mishra, T. V. C. Rao, 
V. C. Sahni and H. Takeya, {\it Proceedings of the DAE Solid State Physics 
Symposium} {\bf 40C} (1997) 310.
\bibitem{bean} C. P. Bean, {\it Phys. Rev. Lett.} { 8} (1962) 250.
\bibitem{reentrant} K. Ghosh, S. Ramakrishnan, A.K. Grover, Gautam I. Menon, 
Girish Chandra, T. V. C. Rao, G. Ravikumar, P. K. Mishra, V. C. Sahni,
C.V. Tomy, G. Balakrishnan, D. Mck. Paul and S. Bhattacharya,
{\it Phys. Rev. Lett.} { 76} (1996) 4600.
\bibitem{satya13} S. S. Banerjee, S. Ramakrishnan, A. K. Grover, G. Ravikumar,
P. K. Mishra, V. C. Sahni, C. V. Tomy, G. Balakrishnan, D. Mck. Paul, 
M. J. Higgins and S. Bhattacharya,
{\it Euro Physics Lett.} 44 (1998) 995.
\bibitem{satya12} S. S. Banerjee, S. Ramakrishnan, A. K. Grover, G. Ravikumar,
P. K. Mishra, V. C. Sahni, C. V. Tomy, G. Balakrishnan, D. Mck. Paul, 
P. L. Gammel, D. J. Bishop, E. Bucher, M. J. Higgins and S. Bhattacharya,
(to appear in Phys. Rev. B); http://xxx.lanl.gov/abs/cond-mat/9907111.   
\bibitem{paltiel0} Y. Paltiel, E. Zeldov, Y. N. Myasoedov,
H. Shtrikman, S. Bhattacharya, M. J. Higgins, Z. L. Xiao, E. Y.
Andrei, P. L. Gammel and D. J. Bishop,(submitted to Phys. Rev. Lett.)
\bibitem{nishizaki} T. Nishizaki, K. Shibata, T. Naito,
M. Maki and N. Kobayashi, {\it J. Low Temp.  Phys.} { 117} (1999) 1375.
\bibitem{ishida} T. Ishida, K. Okuda and H. Asaoka, {\it Phys. Rev. B}
{ 56} (1997) 5128.
\bibitem{shi} J. Shi, X.S. Ling, R. Liang, D. A. Bonn and W. N. Hardy,
{\it Phys. Rev. B} { 60} (1999) R12593.
\bibitem{notabene} The abrupt character\cite{jsps,paltiel0} of the 
low-field transition from the ``Plateau Effect'' regime to the 
``Power Law'' or Bragg glass regime suggests a sharp transition,
supporting the phase diagram of Fig. 1.
\bibitem{nravi} G. Ravikumar, V. C. Sahni, A.K. Grover, S. Ramakrishnan,
P. L. Gammel, D. J. Bishop, E. B\"u"cher, M. J. Higgins and S. Bhattacharya, 
unpublished. 
\bibitem{zhang} In this regard, see, also, J. Zhang, L. E. De Long, 
V. Majidi and R. C. Budhani, {\it Phys. Rev. B} { 53} (1996) R8851.
\bibitem{paltiel} Y. Paltiel, E. Zeldov, Y. N. Myasoedov,
H. Shtrikman, S. Bhattacharya, M. J. Higgins, Z. L. Xiao, E. Y.
Andrei, P. L. Gammel and D. J. Bishop, {\it Nature} 403 (2000) 398.
\bibitem{otterlo} A. van Otterlo, R. T. Scalettar and G. T. Zimanyi, 
{\it Phys.  Rev. Lett.} { 81} (1998) 1497.
\bibitem{deligiannis} K. Deligiannis, P. A. J. de Groot, M. Oussena, 
S. Pinfold, R.  Langan, R. Gagnon and L. Taileffer, {\it Phys. Rev. 
Lett.} { 79} (1997) 2121.
\bibitem{otterlo1} A. van Otterlo, R. T. Scalettar, G. T. Zimanyi,
R. Olsson, A. Petrean, W. Kwok and V. Vinokur, {\it Phys. Rev. Lett.}
{ 84} (2000) 2493.
\bibitem{pal3} D. Pal, D. Dasgupta, Bimal K. Sarma, S. Bhattacharya,
S. Ramakrishnan and A. K. Grover, (to appear in Phys. Rev. B (2000)).
\bibitem{disprl} G. I. Menon and C. Dasgupta, {\it Phys. Rev. Lett.}
{ 73}  (1994) 1023.    
\bibitem{tang} C. Tang, X. S. Ling, Budnick and S. Bhattacharya,
{\it Euro Physics Lett.} 35 (1996) 597.
\bibitem{rudnev} I.A. Rudnev, V. A. Kashurnikov, M. E. Gracheva and
O. A. Nikitenko, {\it Physica C} { 332} (2000) 383.
\end{thebibliography}
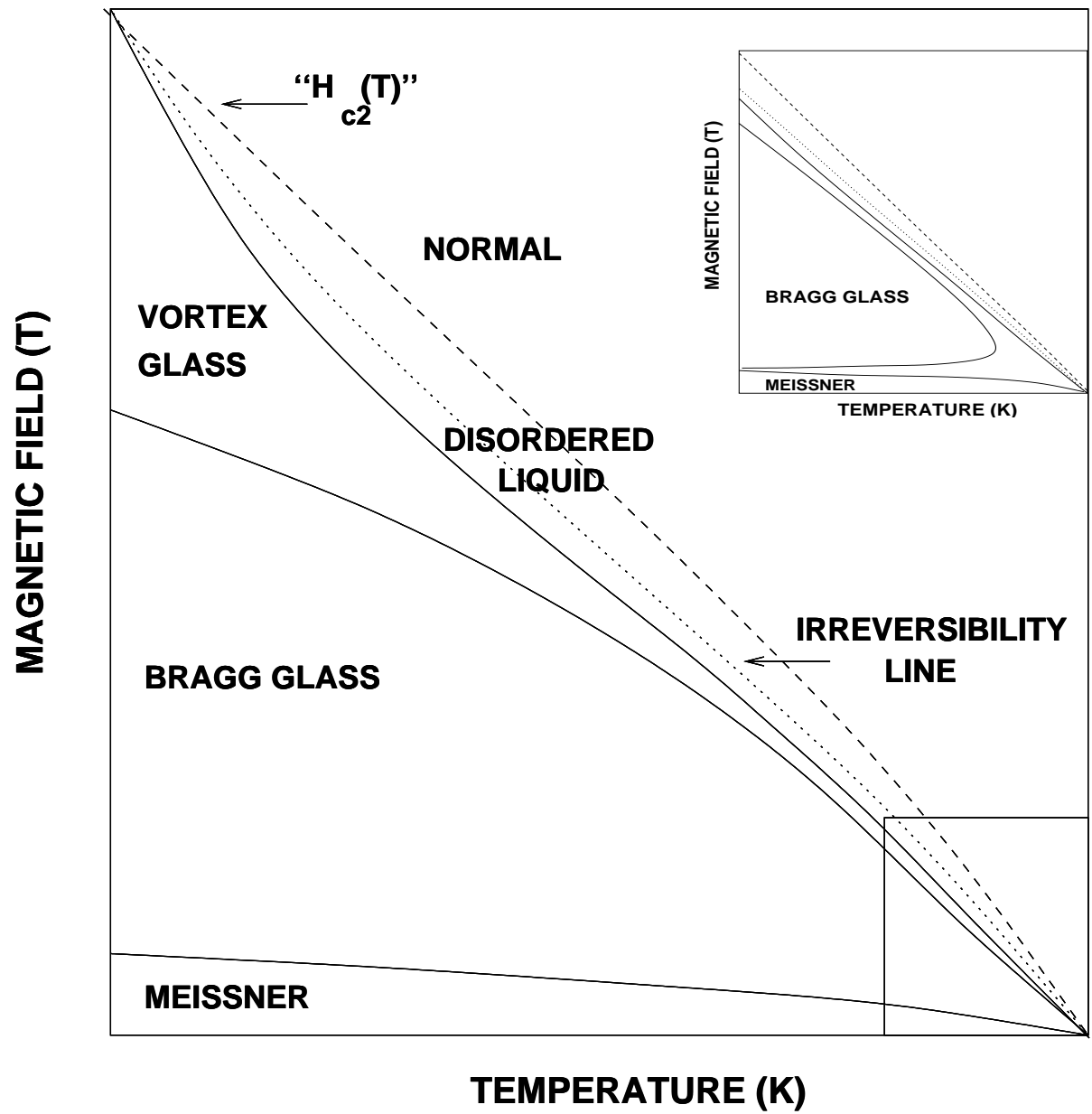
\end{document}